\documentclass{aa}  

\usepackage{graphicx}
\usepackage{amsmath,amsfonts,amssymb}
\usepackage{natbib}

\usepackage{txfonts}
\usepackage{xcolor}

\usepackage{blindtext}
\usepackage{float}
\usepackage{dblfloatfix}
\usepackage{afterpage}
\usepackage{ifthen}
\usepackage[morefloats=12]{morefloats}

\usepackage{placeins}
\usepackage{multicol}
\bibpunct{(}{)}{;}{a}{}{,}
\usepackage[switch]{lineno}
\definecolor{linkcolor}{rgb}{0.6,0,0}
\definecolor{citecolor}{rgb}{0,0,0.75}
\definecolor{urlcolor}{rgb}{0.12,0.46,0.7}
\usepackage[breaklinks, colorlinks, urlcolor=urlcolor,
    linkcolor=linkcolor,citecolor=citecolor,pdfencoding=auto]{hyperref}
\hypersetup{linktocpage}
\usepackage{bold-extra}

\def\setsymbol#1#2{\expandafter\def\csname #1\endcsname{#2}}
\def\getsymbol#1{\csname #1\endcsname}

\def\Planck{\textit{Planck}}

\newbox\tablebox    \newdimen\tablewidth
\def\leaderfil{\leaders\hbox to 5pt{\hss.\hss}\hfil}
\def\tablenote#1 #2\par{\begingroup \parindent=0.8em
    \abovedisplayshortskip=0pt\belowdisplayshortskip=0pt
    \noindent
    $$\hss\vbox{\hsize\tablewidth \hangindent=\parindent \hangafter=1 \noindent
    \hbox to \parindent{$^#1$\hss}\strut#2\strut\par}\hss$$
    \endgroup}

\def\L2{\ifmmode L_2\else $L_2$\fi}

\def\DeltaT{\ifmmode \Delta T\else $\Delta T$\fi}
\def\deltat{\ifmmode \Delta t\else $\Delta t$\fi}
\def\fknee{\ifmmode f_{\rm knee}\else $f_{\rm knee}$\fi}
\def\Fmax{\ifmmode F_{\rm max}\else $F_{\rm max}$\fi}
\def\solar{\ifmmode{\rm M}_{\mathord\odot}\else${\rm M}_{\mathord\odot}$\fi}
\def\Msolar{\ifmmode{\rm M}_{\mathord\odot}\else${\rm M}_{\mathord\odot}$\fi}
\def\Lsolar{\ifmmode{\rm L}_{\mathord\odot}\else${\rm L}_{\mathord\odot}$\fi}
\def\inv{\ifmmode^{-1}\else$^{-1}$\fi}
\def\mo{\ifmmode^{-1}\else$^{-1}$\fi}
\def\sup#1{\ifmmode ^{\rm #1}\else $^{\rm #1}$\fi}
\def\expo#1{\ifmmode \times 10^{#1}\else $\times 10^{#1}$\fi}
\def\,{\thinspace}
\def\lsim{\mathrel{\raise .4ex\hbox{\rlap{$<$}\lower 1.2ex\hbox{$\sim$}}}}
\def\gsim{\mathrel{\raise .4ex\hbox{\rlap{$>$}\lower 1.2ex\hbox{$\sim$}}}}

\def\simprop{\mathrel{\raise .4ex\hbox{\rlap{$\propto$}\lower 1.2ex\hbox{$\sim$}}}}
\def\deg{\ifmmode^\circ\else$^\circ$\fi}
\def\pdeg{\ifmmode $\setbox0=\hbox{$^{\circ}$}\rlap{\hskip.11\wd0 .}$^{\circ}
          \else \setbox0=\hbox{$^{\circ}$}\rlap{\hskip.11\wd0 .}$^{\circ}$\fi}
\def\arcs{\ifmmode {^{\scriptstyle\prime\prime}}
          \else $^{\scriptstyle\prime\prime}$\fi}
\def\arcm{\ifmmode {^{\scriptstyle\prime}}
          \else $^{\scriptstyle\prime}$\fi}
\newdimen\sa  \newdimen\sb
\def\parcs{\sa=.07em \sb=.03em
     \ifmmode \hbox{\rlap{.}}^{\scriptstyle\prime\kern -\sb\prime}\hbox{\kern -\sa}
     \else \rlap{.}$^{\scriptstyle\prime\kern -\sb\prime}$\kern -\sa\fi}
\def\parcm{\sa=.08em \sb=.03em
     \ifmmode \hbox{\rlap{.}\kern\sa}^{\scriptstyle\prime}\hbox{\kern-\sb}
     \else \rlap{.}\kern\sa$^{\scriptstyle\prime}$\kern-\sb\fi}
\def\ra[#1 #2 #3.#4]{#1\sup{h}#2\sup{m}#3\sup{s}\llap.#4}
\def\dec[#1 #2 #3.#4]{#1\deg#2\arcm#3\arcs\llap.#4}
\def\deco[#1 #2 #3]{#1\deg#2\arcm#3\arcs}
\def\rra[#1 #2]{#1\sup{h}#2\sup{m}}

\def\dots{\relax\ifmmode \ldots\else $\ldots$\fi}
\def\WHzsr{\ifmmode $W\,Hz\mo\,sr\mo$\else W\,Hz\mo\,sr\mo\fi}
\def\mHz{\ifmmode $\,mHz$\else \,mHz\fi}
\def\GHz{\ifmmode $\,GHz$\else \,GHz\fi}
\def\mKs{\ifmmode $\,mK\,s$^{1/2}\else \,mK\,s$^{1/2}$\fi}
\def\muKs{\ifmmode \,\mu$K\,s$^{1/2}\else \,$\mu$K\,s$^{1/2}$\fi}
\def\muKRJs{\ifmmode \,\mu$K$_{\rm RJ}$\,s$^{1/2}\else \,$\mu$K$_{\rm RJ}$\,s$^{1/2}$\fi}
\def\muKHz{\ifmmode \,\mu$K\,Hz$^{-1/2}\else \,$\mu$K\,Hz$^{-1/2}$\fi}
\def\MJysr{\ifmmode \,$MJy\,sr\mo$\else \,MJy\,sr\mo\fi}
\def\MJysrmK{\ifmmode \,$MJy\,sr\mo$\,mK$_{\rm CMB}\mo\else \,MJy\,sr\mo\,mK$_{\rm CMB}\mo$\fi}
\def\microns{\ifmmode \,\mu$m$\else \,$\mu$m\fi}

\def\muK{\ifmmode \,\mu$K$\else \,$\mu$\hbox{K}\fi}
\def\microK{\ifmmode \,\mu$K$\else \,$\mu$\hbox{K}\fi}
\def\muW{\ifmmode \,\mu$W$\else \,$\mu$\hbox{W}\fi}
\def\kms{\ifmmode $\,km\,s$^{-1}\else \,km\,s$^{-1}$\fi}
\def\kmsMpc{\ifmmode $\,\kms\,Mpc\mo$\else \,\kms\,Mpc\mo\fi}
\providecommand{\sorthelp}[1]{}

\def\Cosmoglobe{\textsc{Cosmoglobe}}
\def\Planck{\textit{Planck}}

\begin{document}

   \title{\Cosmoglobe: Simulating zodiacal emission with ZodiPy \thanks{This work is part of the Cosmoglobe effort, and the code is published under an open-source license at \url{https://github.com/Cosmoglobe/zodipy}.}}

   \author{M.~San, D.~Herman, G. B.~Erikstad, M.~Galloway, and D.~Watts}

   \institute{Institute of Theoretical Astrophysics, University of Oslo, Blindern, Oslo, Norway}
  
   \titlerunning{\Cosmoglobe: Simulating zodiacal emission with ZodiPy}
   \authorrunning{M.~San et al.}

   \date{\today}
   
  \abstract{We present ZodiPy, a modern and easy-to-use Python package for modeling the zodiacal emission seen by an arbitrary Solar System observer, which can be used for the removal of both thermal emission and scattered sunlight from interplanetary dust in astrophysical data. The code implements the \textrm{COBE} Diffuse Infrared Background Experiment (DIRBE) interplanetary dust model and the \Planck\ extension, which allows for zodiacal emission predictions at infrared wavelengths in the 1.25--240 $\mu$m range and at microwave frequencies in the 30--857 GHz range. The predicted zodiacal emission may be extrapolated to frequencies and wavelengths not covered by the built-in models to produce forecasts for future experiments. ZodiPy attempts to enable the development of new interplanetary dust models by providing the community with an easy-to-use interface for testing both current and future models.
  We demonstrate how the software can be used by creating simulated zodiacal emission timestreams for the DIRBE experiment and show that these agree with corresponding timestreams produced with the DIRBE Zodiacal Light Prediction Software. We also make binned maps of the zodiacal emission as predicted to be observed by DIRBE and compare them with the DIRBE calibrated individual observations (CIO).}


   \maketitle

\section{Introduction}

Zodiacal emission (ZE, sometimes called zodiacal light emission or interplanetary dust emission) is a source of radiation observable from the optical to the submillimeter regimes (\citealt{LEINERT1997} and references therein). Observations of ZE date back millennia and still hold prominence for a host of modern experiments. 
ZE is caused by scattering and re-emission at infrared wavelengths of sunlight absorbed by interplanetary dust (IPD) within the Solar System. The diffuse ZE is a major contributor to the total sky brightness at infrared wavelengths between 1--100 $\mu$m, and it dominates the mid-infrared ($\sim 10$--60~$\mu$m) with a spectral peak at around 10--20~$\mu$m.

Zodiacal emission has been a driving force in the exploration of the interplanetary medium since the seventeenth century \citep{CASSINI}, and it was found to be well characterizable in the infrared by the Diffuse Infrared
Background Experiment (DIRBE) instrument onboard the Cosmic Background Explorer (\textrm{COBE}) \citep{mather:1994, hauser:1998}. 
With the purpose of removing ZE from their data, the DIRBE team constructed a geometric model depicting the interplanetary medium and its distinguishable components. These components were then set up with modified blackbody spectra and evaluated through line-of-sight integration to make predictions of the observed ZE. This model, described in \cite{K98} (hereafter K98), has proven to be effective for describing the ZE in the infrared and submillimeter regimes. The DIRBE model was recently used by the \Planck\ Collaboration \citep{PLANCK, Planck2015, Planck2018} in their analysis of the High Frequency Instrument (HFI) data, where it was adapted to be valid at subterahertz frequencies. The \textit{Planck} update to the DIRBE model is the state-of-the-art method for removing ZE from data in cosmic microwave background (CMB) cosmology.

In CMB studies, the ZE is a local foreground whose structure is highly dependent on the position of the observer within the Solar System. A consequence of this is that every experiment observes a unique ZE signal due to differences in the scanning strategy and telescope position. As such, it is impossible to describe the ZE foreground through a single template, as would be possible for most Galactic foregrounds, and instead ZE must be dynamically modeled on a per-experiment basis.

While new data will refine our understanding of the interplanetary medium, it is necessary for future models to be consistent with both new and archival data. It is precisely this need that motivates the \textsc{Cosmoglobe} project,\footnote{Learn more about \Cosmoglobe\ at \newline\url{https://www.cosmoglobe.uio.no/}} which aims to create a framework that will allow for the refinement of astrophysical models jointly with the raw data from complementary experiments. This form of joint analysis is already being explored within the framework of \textrm{WMAP} \citep{bp17} and \textrm{LiteBIRD} \citep{bp16}, in combination with \textit{Planck} Low Frequency Instrument data \citep{bp01}.
ZE is an especially promising direction for joint analysis, in part because HFI made observations of ZE at complementary wavelengths to DIRBE. While the DIRBE model was modified during the HFI analysis, no attempt was made to improve upon the geometrical representation of the model components using the larger effective dataset. Our understanding of the interplanetary medium has improved since the development of the DIRBE IPD model  (see, for example, \citealp{REACHBANDS, REACHRING, REACHFEATURE}). The AKARI satellite, which observed in the infrared at wavelengths between 6--180 $\mu$m, detected small-scale structures in the ZE which were not well-characterized by the DIRBE model \citep{AKARI}. While the DIRBE model has been successful in describing the large-scale diffuse ZE, modern high-resolution high-frequency and infrared experiments will require IPD models that can more effectively resolve the small-scale structures in the ZE. As such, an update of the community state-of-art ZE model is long overdue. In this \textsc{Cosmoglobe} framework, such a model refinement is a natural byproduct of a joint analysis of DIRBE, HFI, and other data. To improve upon this model, it is essential to have the model agree with the data at all wavelengths. The modeling of ZE is in the process of being implemented in the \texttt{Commander} framework, which will ultimately be used for the joint processing of HFI and DIRBE data. However, a stand-alone code is useful for agile model building and data analysis. As such, ZodiPy will function as the first step toward this goal in the \textsc{Cosmoglobe} framework, in addition to making ZE corrections on arbitrary data more accessible to the community.

In this paper we present ZodiPy, a modern Python implementation of the DIRBE model in a user-friendly fashion which allows for forecasting of ZE by arbitrary Solar System observers at wavelengths of 1.25--240 $\mu$m and frequencies of 30--857 GHz. ZodiPy supports custom user-defined model implementations, and it allows for extrapolations of models to wavelengths or frequencies outside of the valid ranges defined by the models. In addition to forecasting ZE for current and future experiments, ZodiPy will act as a stand-alone code for testing and comparing improved models of ZE with the \Cosmoglobe\ framework.

In Sect. \ref{sec: ipd model} we introduce the DIRBE IPD model and show geometric illustrations of its IPD components. In Sect. \ref{sec: zodiemission} we introduce the emission mechanisms of IPD and describe the implementation of the model evaluation in ZodiPy. Additionally, we present instantaneous maps of the ZE, both the total emission and the component-wise emission. In Sect. \ref{sec: DIRBE} we illustrate a few use-cases of ZodiPy by applying the software to make predictions about the ZE foreground as observed by DIRBE.

\section{Interplanetary dust model}\label{sec: ipd model}
ZodiPy implements the DIRBE IPD model as described in \cite{K98}. The model is twofold. The first part is a three-dimensional parametric description of the distribution of IPD in the Solar System, consisting of six individual components: the diffuse cloud, three dust bands, the circumsolar ring, and the Earth-trailing feature. A detailed description of each component is presented in Sec. \ref{sec: ipd comps}. The second part of the model describes the emission mechanisms of IPD. In total, there are over 80 free parameters in the DIRBE model. In the following sections, we introduce the parameterized number density functions of the IPD components.


\subsection{Coordinates and geometry}
All IPD components, except the Earth-trailing feature, are distributed symmetrically around the Sun with respect to the ecliptic plane. The natural coordinate system for describing the three-dimensional density distribution of an IPD component, $c$, is Cartesian heliocentric ecliptic coordinates $(x, y, z)$. However, the spatial origin of the components is not necessarily aligned perfectly with that of the Sun. Therefore a coordinate offset is introduced for each component $(x_{0,c}, y_{0,c}, z_{0,c})$. We define these component-centric coordinates as follows:
\begin{equation}
    \begin{aligned}
    x_c&= x - x_{0,c}\\
    y_c&= y - y_{0,c}\\
    z_c&= z - z_{0,c}.
    \end{aligned}
\end{equation}
In addition to this offset, each component is allowed to have some inclination $i$ and ascending node $\Omega$, with respect to the ecliptic. The symmetry of a component with respect to its mid-plane allows us to completely describe its density with the two coordinates
\begin{align}
R_c &= \sqrt{x_c^2 + y_c^2 + z_c^2}\\
Z_c &= x_c\sin{\Omega_c}\sin{i_c} - y_c \cos{\Omega_c}\sin{i_c} + z_c \cos{i_c},
\end{align}
where $R_c$ is the radial distance from the center of the component, and $Z_c$ is the height above its mid-plane. The ratio between these two coordinates, $\zeta_c \equiv |Z_c / R_c|$, is a helpful auxiliary quantity for describing the fan-like distribution that the diffuse cloud and the dust bands exhibit. We note that all distances in the DIRBE IPD model are given in units of AU and implicitly divided by 1~AU.

\subsection{Interplanetary dust components}\label{sec: ipd comps}
Below we introduce the DIRBE IPD components. We refer the reader to K98 (and references therein) for an in-depth review of the observational and physical motivations behind the model components.

\subsubsection{The diffuse cloud}
\begin{figure*}
  \centering
   	\includegraphics[width=\linewidth]{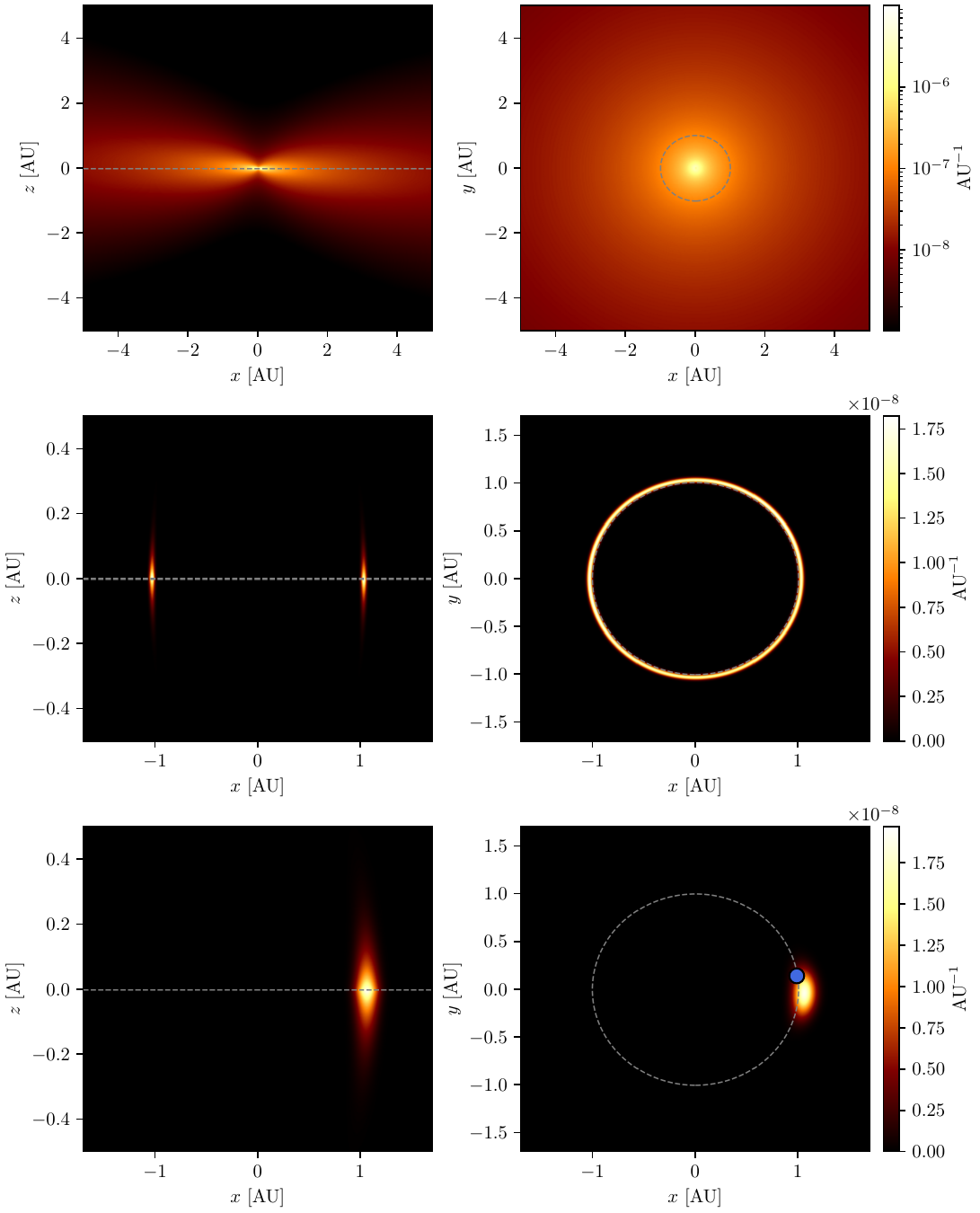}
  	\caption{Interplanetary dust density distribution of the diffuse cloud, the circumsolar ring, and the Earth-trailing feature. (\textit{Left column:}) Cross section showing the $xz$ plane ($y=0$). The gray-dashed line represents the ecliptic plane. (\textit{Right column:}) Cross section showing the $xy$ plane ($z=0$). The gray-dashed circle represents the orbit of Earth. (\textit{Top row:}) The density of the diffuse cloud. (\textit{Middle row:}) The density of the circumsolar ring. (\textit{Bottom row:}) The density of the Earth-trailing feature. The blue dot represents Earth.}
	\label{fig: denscloudringfeature}
\end{figure*}

The primary component in the model is the diffuse cloud. As the name suggests, this is a cloud-like component that encapsulates the inner Solar System and accounts for the majority of total IPD density in the model. The number density of the diffuse cloud, $n_\mathrm{C}$, is modeled in a separable form with a radial and vertical component,
\begin{equation}
n_\mathrm{C}(R_\mathrm{C}, Z_\mathrm{C}) = n_{0,\mathrm{C}} R_\mathrm{C}^{-\alpha} f(\zeta_\mathrm{C}).
\end{equation}
The radial term, $R_\mathrm{C}^{-\alpha}$ is a power law with $\alpha$ as a free parameter. The vertical term, $f(\zeta_\mathrm{C})$, represents a widened modified fan model,
\begin{equation}
    f(\zeta)=\exp[-\beta g^\gamma],
\end{equation}
where
\begin{equation}
g=\left\{\begin{array}{cl}
\zeta^{2} / 2 \mu & \text { for } \zeta<\mu \\
\zeta-\mu / 2 & \text { for } \zeta \geq \mu,
\end{array}\right.
\end{equation}
and $\beta$, $\gamma$, and $\mu$ are free parameters describing the vertical shape and widening of the diffuse cloud. The density distribution of the diffuse cloud, as detailed by the DIRBE model, is illustrated in the top panel of Fig. \ref{fig: denscloudringfeature}, where we show two cross sections of the tabulated cloud density in the $xz$ and $xy$ planes, respectively.

\subsubsection{Dust bands}
\begin{figure*}
   	\includegraphics[width=0.999\linewidth]{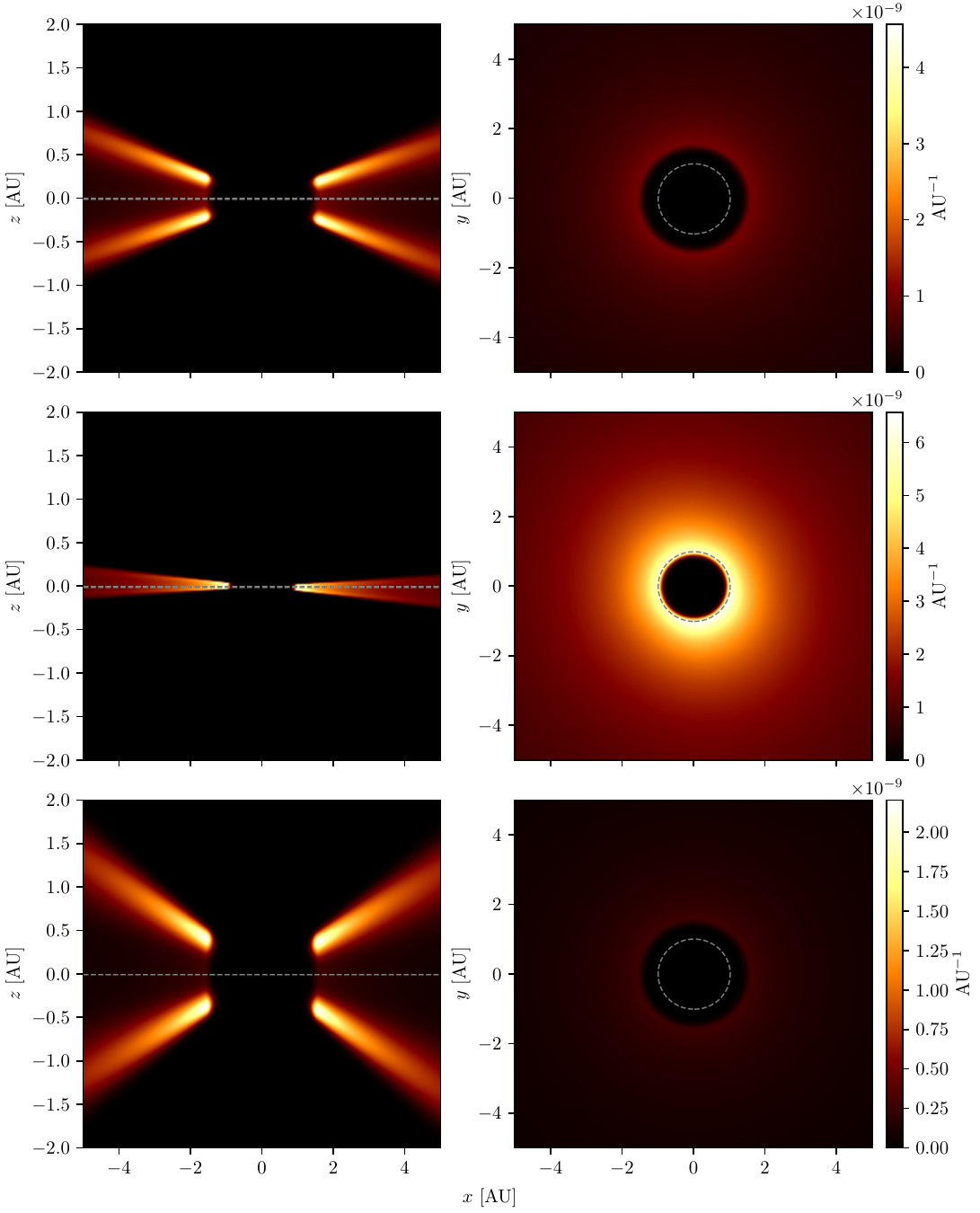}
  	\caption{Interplanetary dust density distribution of the three dust bands. (\textit{Left column:}) Cross section showing the $xz$ plane ($y=0$). The gray-dashed line represents the ecliptic plane. (\textit{Right column:}) Cross section showing the $xy$ plane ($z=0$). The gray-dashed circle represents the orbit of Earth. (\textit{Top row:}) The density of dust band 1. (\textit{Middle row:}) The density of dust band 2. (\textit{Bottom row:}) The density of dust band 3.}
  	\label{fig: densbands}
\end{figure*}
The model includes three dust band pairs. These are bands of IPD concentrations appearing at distinct ecliptic latitudes associated with asteroid families \citep{sykes1990}. The density of a dust band, $\mathrm{B}_j$, where $j \in\{1,2,3\}$, is modeled in the following form:
\begin{equation}
    \begin{aligned}
        n_{\mathrm{B}_j}(R_{\mathrm{B}_j}, Z_{\mathrm{B}_j})=& \frac{3 n_{0, \mathrm{B}_j}}{R_{\mathrm{B}_j}} \exp \left[-\left(\frac{\zeta_{\mathrm{B}_j}}{\delta_{\zeta_{\mathrm B_j}}}\right)^{6}\right]\left[1 + \left(\frac{\zeta_{\mathrm{B}_j}}{\delta_{\zeta_{\mathrm{B}_j}}}\right)^{p_{\mathrm{B}_j}}v^{-1}_{\mathrm{B}_j}\right] \\
        & \times\left\{1-\exp \left[-\left(\frac{R_{\mathrm{B}_j}}{\delta_{R_{\mathrm{B}_j}}}\right)^{20}\right]\right\},
    \end{aligned}
\end{equation}
where $n_{0, {\mathrm{B}_j}}$ is the density of the band at 3 AU, $\delta_{R_{\mathrm{B}_j}}$ is a free parameter representing the inner radial cutoff, and $\delta_{\zeta_{\mathrm{B}_j}}$, $v_{\mathrm{B}_j}$, and $p_{\mathrm{B}_j}$ are free shape parameters. We note that there is a factor of $v_{\mathrm{B}_j}$ difference between this formulation and Eq.~(8) in K98. This factor is mentioned in Sect.~4.1.2 in \cite{PLANCK} and comes from a difference between the text and the implementation in the \textrm{COBE} DIRBE Zodiacal Light Prediction Software.\footnote{The \textrm{COBE} DIRBE Zodiacal Light Prediction Software is a code written in IDL which computes the ZE predicted to be observed by DIRBE. The source code is available on LAMBDA: \url{https://lambda.gsfc.nasa.gov/product/cobe/dirbe_zodi_sw.html}.} We follow the convention in the code implementation. The three band pairs, which appear at ecliptic latitudes around $\pm 1.4^\circ$, $\pm 10^\circ$, and $\pm 15^\circ$, respectively, are illustrated in Fig. \ref{fig: densbands}, where we show two cross sections of the tabulated band densities in the $xz$ and $xy$ planes, respectively.

\label{sec: ringfeature}
\subsubsection{Circumsolar ring and Earth-trailing feature}
As the orbit of IPD grains decays, they may end up getting trapped in gravitational resonant orbits with planets, resulting in circumsolar rings. The DIRBE model includes one such circumsolar ring, $n_\mathrm{R}$, for Earth's orbit,
\begin{equation}\label{eq: ring}
    n_\mathrm{R}(R_\mathrm{R}, Z_\mathrm{R})=n_{\mathrm{R},0} \exp \left[-\frac{\left(R_\mathrm{R}-R_{0, \mathrm{R}}\right)^2}{\sigma_{\mathrm{R}, r} ^2}-\frac{\left| Z_\mathrm{R} \right|}{\sigma_{\mathrm{R}, z}}\right],
\end{equation}
where $n_\mathrm{R}$ is the density of the ring at 1 AU, $R_{0, \mathrm{R}}$ is the radius of the peak density, and $\sigma_{\mathrm{R}, r}$ and $\sigma_{\mathrm{R}, z}$ are the radial and vertical dispersions, respectively.

\cite{DERMOTT1994} showed through dynamical simulations of the circumsolar ring that we should expect enhancements in the dust concentration of the ring in the regions trailing and leading Earth. The DIRBE team was able to fit such a density feature, $n_\mathrm{F}$, to the region trailing Earth,
\begin{equation}\label{eq: feature}
   n_\mathrm{F}(R_\mathrm{F}, Z_\mathrm{F}, \theta_\mathrm{F}) = n_{\mathrm{F}, 0} \exp \left[-\frac{\left(R_\mathrm{F}-R_{\mathrm{F}, 0}\right)^{2}}{\sigma_{\mathrm{F}, r}^{2}}-\frac{\left|Z_\mathrm{F}\right|}{\sigma_{\mathrm{F}, z}}-\frac{\left(\theta_\mathrm{F}-\theta_{\mathrm{F}, 0}\right)^{2}}{\sigma_{\mathrm{F}, \theta }^{2}}\right],
\end{equation}
where $\theta_\mathrm{F}$ is the ecliptic longitude of Earth seen from the feature-centric frame. The Earth-trailing feature is the only component in the model not symmetrical in its mid-plane, which will require us knowing the heliocentric longitude of Earth at all observation times to determine the position of the Earth-trailing feature. We note that there is a factor $1/2$ difference in the first term in the exponential of Eq. \eqref{eq: ring}, and the first and last term in the exponential of Eq. \eqref{eq: feature} and Eq.~(9) in K98, which is again due to a difference between the text and the code implementation. The density distributions of the circumsolar ring and the Earth-trailing feature are illustrated in the middle and bottom panels of Fig. \ref{fig: denscloudringfeature}, where we show two cross sections of the tabulated densities in the $xz$ and $xy$ planes, respectively.

\section{Zodiacal emission}\label{sec: zodiemission}
Now that we have a model for the spatial distribution of IPD in the Solar System, the next step is to connect the densities of each IPD component to the foreground emission seen by a Solar System observer. In the following sections, we illustrate the characteristic time variation of the ZE and give an overview of the emission mechanisms of IPD as described in K98. We also show how we apply the full DIRBE model to evaluate the ZE in ZodiPy.

\subsection{Time-varying emission}\label{sec: time-varying}
\begin{figure}
  \centering
	\includegraphics[width=0.7\linewidth]{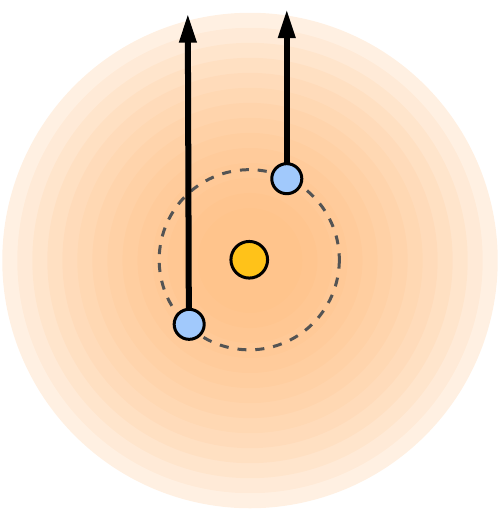}
	\caption{Illustration showing that the integrated IPD along a line of sight toward a point on the celestial sphere as seen from Earth (blue circles) changes as Earth orbits the Sun (yellow circle).}
	\label{fig: illustration}
\end{figure}
The motion of an observer through the IPD distribution introduces a time dependence in the observed ZE. To an observer orbiting the Sun with Earth, the IPD distribution moves along the ecliptic by approximately one degree a day (with respect to the Galactic coordinate frame). As the observer travels within this distribution, any line of sight toward the same point on the distant celestial sphere contains varying amounts of IPD. This effect is illustrated in Fig. \ref{fig: illustration}. 

The characteristic time variation in ZE makes it easy to recognize in data, even at wavelengths and frequencies where the observed emission is very weak. We can extract the ZE signal by splitting the data temporally (with similar sky coverage) and subtracting one from the other, which removes any time-invariant Galactic emission. On the other hand, a downside of this time variation is that the emission becomes a function of the scanning strategy of experiments, making it impossible to construct a universal foreground template of the ZE, as is commonly done for Galactic foregrounds. Instead, the ZE observed by an experiment must be dynamically modeled on a per-experiment basis.

\subsection{Emission mechanisms}\label{sec: emission}
When heated by their surroundings, dust grains emit thermal radiation. In the case of IPD grains, the heat source is radiation from the Sun. If we assume that each dust grain behaves similar to a blackbody, then the emission from a single dust grain at some delta wavelength, $\lambda$, is given by the Planck function $B_\lambda(T)$,
\begin{equation}
    I^\mathrm{Thermal}_{\lambda} = B_\lambda(T).
\end{equation}
The temperature of the dust grain, $T$, is assumed to be a function of the radial distance from the Sun,
\begin{equation}
T(R) = T_0 R^{-\delta},
\end{equation}
where $T_0$ is the temperature of the dust grain at 1 AU, $R$ is the distance from the Sun, and $\delta$ is the power-law exponent describing how the temperature falls with distance from the Sun.

Dust grains come in varying shapes and compositions and they are not true blackbodies. We, therefore, model the IPD grains as modified blackbodies instead, where we scale the Planck function, $B_\lambda(T)$, by an emissivity, $E_\lambda$, which measures the deviation of the IPD grain from the Planck function at some wavelength. Additionally, we assume that all the dust in a specific IPD component is of the same composition, but we allow the composition to vary from component to component, such that the emissivity becomes $E_{c,\lambda}$. The thermal emission then becomes
\begin{equation}
    I^\mathrm{Thermal}_{c,\lambda} = E_{c,\lambda} B_\lambda(T(R)).
\end{equation}

In addition to generating thermal radiation, IPD grains also scatter sunlight at infrared wavelengths. The scattering term, as defined in K98, is 
\begin{equation}\label{eq: scat_term}
    I^\mathrm{Scattering}_{c, \lambda} = A_{c, \lambda} F_\lambda^\odot(R) \Phi_\lambda(\Theta).
\end{equation}
Here $A_{c, \lambda}$ is the albedo, which represents the fraction of photons that get reflected by the surface of the grain and is a function of the grain composition, and $F_\lambda^\odot$ is the solar spectral irradiance, representing the flux received at some wavelength by the surface of the dust grains at some distance from the Sun.  The spectral solar irradiance is given by
\begin{equation}
    F_\lambda^\odot(R) = \frac{F_{\lambda,0}^\odot}{R^2},
\end{equation}
where $F_{\lambda,0}^\odot$ is the reference spectra of the solar irradiance at 1~AU, and $R$ is the radial distance from the Sun. The solar irradiance reference spectra used in the DIRBE model is tabulated in the \textrm{COBE} DIRBE Zodiacal Light Prediction Software.

Finally, $\Phi_\lambda(\Theta)$ is the phase function and represents the distribution of scattered light intensity. The phase function as defined in K98 is
\begin{equation}
    \Phi_{\lambda}(\Theta)=N\left[C_{0, \lambda}+C_
    {1, \lambda} \Theta+\mathrm{e}^{C_{2, \lambda} \Theta}\right],
\end{equation}
where $C_{k, \lambda}$ are free parameters.
The phase function can be interpreted as a probability density, representing the probability of a photon getting scattered as a function of the scattering angle $\Theta$, and it must therefore be normalized. The DIRBE normalization factor $N$ has the following analytic solution:
\begin{equation}
    N = \left[ 2\pi \left( 2 C_{0, \lambda} + \pi C_{1, \lambda} + \frac{\mathrm{e}^{\pi C_{2, \lambda}} + 1}{{C^2_{2, \lambda}} + 1} \right)\right]^{-1}.
\end{equation}
At scattering wavelengths, that is to say wavelengths where $A_{c, \lambda} \ne 0$, we assume some fraction of the total radiated heat to be scattered away from the observer, such that the thermal contribution becomes
\begin{equation}\label{eq: therm_term}
    I^\mathrm{Thermal}_{c,\lambda} = \left( 1 - A_{c, \lambda} \right) E_{c,\lambda} B_\lambda(T(R)).
\end{equation}
Combining Eqs. \eqref{eq: scat_term} and \eqref{eq: therm_term} gives us the total emission in intensity from a single grain of IPD,
\begin{align}
    I^\mathrm{Total}_{c, \lambda} &= I^\mathrm{Scattering}_{c,\lambda} + I^\mathrm{Thermal}_{c,\lambda}\\
    &= A_{c, \lambda} F_\lambda^\odot(R) \Phi_\lambda(\Theta) + \left( 1 - A_{c, \lambda} \right) E_{c,\lambda} B_\lambda(T(R)).
\end{align}

\subsection{Evaluating the zodiacal emission}
Now that we have expressions for both the number densities of the IPD components and the emission mechanisms, we can turn to the full model evaluation in ZodiPy. For an experiment observing some point $p$ on the sky at time $t$, the total contribution to the sky intensity from ZE is 
\begin{equation}\label{eq: intensity}
    I_{p,t} = \sum_c \int n_c \left[  A_{c, \lambda} F_\lambda^\odot \Phi_\lambda + \left( 1 - A_{c, \lambda} \right) E_{c,\lambda} B_\lambda \right]\,\mathrm ds,
\end{equation}
where we integrate along a line-of-sight $\mathrm ds$ from the observer toward $p$. Equation~\eqref{eq: intensity} represents the brightness integral as described in Eq.~(1) in K98, but without the color-correction factor, and it is further reduced to Eq.~(8) in \cite{PLANCK} at wavelengths where $A_{c, \lambda}=0$. We note that in the DIRBE analysis, the endpoints of all considered line of sights were fixed to a radial distance of 5.2~AU from the Sun, corresponding to the orbital distance of Jupiter. We would prefer to relax this assumption, but it is likely that the DIRBE IPD model parameters are affected by this cutoff. As such, ZodiPy follows this convention by default, but the user may specify an alternative cutoff distance. The line-of-sight integration is performed using Gaussian-Legendre quadrature, with a default of 100 points, but the user may change this number depending on the desired accuracy.

The pointing in Eq.~\eqref{eq: intensity} may be given as a single sky coordinate or a sequence of sky coordinates, either in the form of angles on the sky ($\theta$, $\phi$) or as integers referring to the indices of pixels on a HEALPix\footnote{HEALPix is an acronym for Hierarchical Equal Area isoLatitude Pixelation of a sphere. For more information on HEALPix, see\newline\url{http://healpix.sf.net}.} pixelization grid with the resolution given by $N_\mathrm{side}$ \citep{HEALPIX}. The pointing information is internally converted to ecliptic unit vectors on the sphere using the \texttt{healpy} software \citep{HEALPY} and further used (alongside the observer position and the radial cutoff) to construct the evaluated line of sights. The observer position is assumed to be constant over the duration of the user-provided input pointing sequence, which means that the pointing must be appropriately chunked depending on the position and velocity of the observer. For instance, observers orbiting the Sun alongside Earth should provide pointing sequences of periods corresponding to a day of observation at most, so as to account for the time variations in the ZE. The predicted ZE corresponding to a pointing sequence may be outputted as either a timestream or as a binned HEALPix map, both in units of $\mathrm{MJy\,sr^{-1}}$. For code examples of how to use ZodiPy, we redirect the reader to the documentation\footnote{The documentation along with example code can be found at\newline\url{https://zodipy.readthedocs.io}.}.

\subsection{Instantaneous emission maps}
\begin{figure*}
    \centering
    \includegraphics[width=0.495\linewidth]{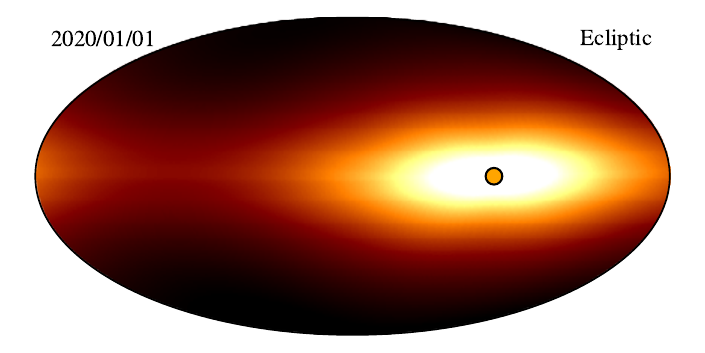}
    \includegraphics[width=0.495\linewidth]{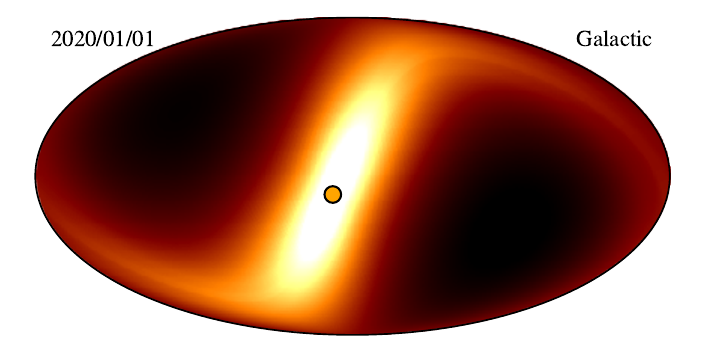}\\ \includegraphics[width=0.495\linewidth]{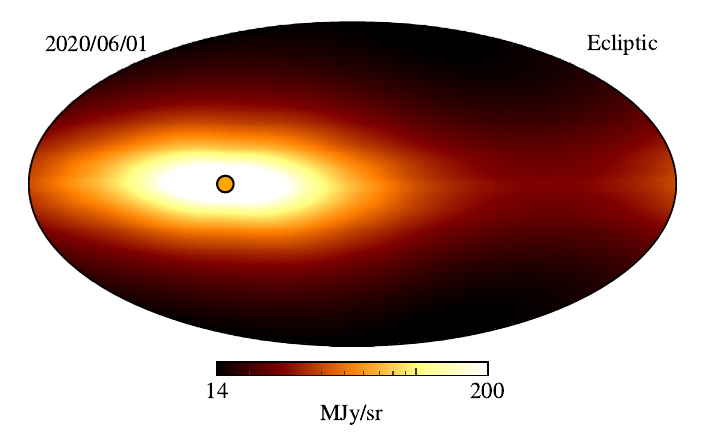}
    \includegraphics[width=0.495\linewidth]{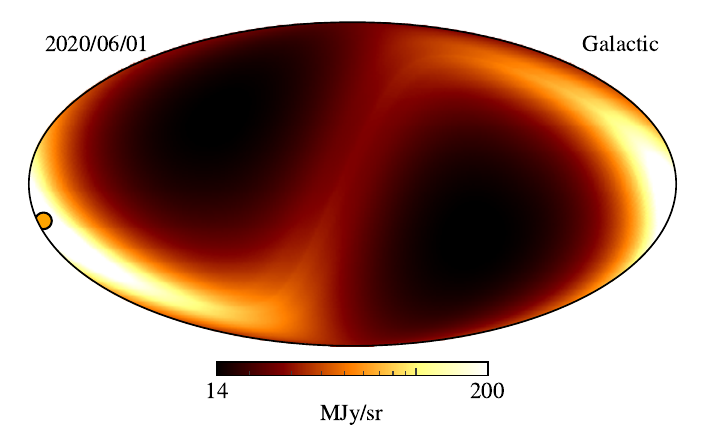}\\
    \caption{Instantaneous maps of the predicted ZE at $25\mu$m as seen by an observer on Earth on January 1, 2020 (\textit{top panel}) and June 1, 2020 (\textit{bottom panel}), in ecliptic (\textit{left column}) and Galactic (\textit{right column}) coordinates. The location of the Sun is marked as an orange circle on all four maps.}
    \label{fig: inst_ecl_gal}
\end{figure*}
In this section, we explore the structure of the ZE as seen on a particular day with the DIRBE model. We make maps of the full-sky ZE in both ecliptic and Galactic coordinates and full-sky maps in ecliptic coordinates of each IPD component, respectively. These maps illustrate how the parametric DIRBE model relates to the observed ZE and functions as a validation of the model implementation. Figure~\ref{fig: inst_ecl_gal} shows the instantaneous emission at 25~$\mu$m seen by an observer on Earth at two different times (January 1, 2020, in the top panel and June 1, 2020, in the bottom panel). These are maps that represent the ZE as seen by an observer able to simultaneously view the full sky at one instant in time.

\begin{figure*}
  \centering
	\includegraphics[width=0.495\linewidth]{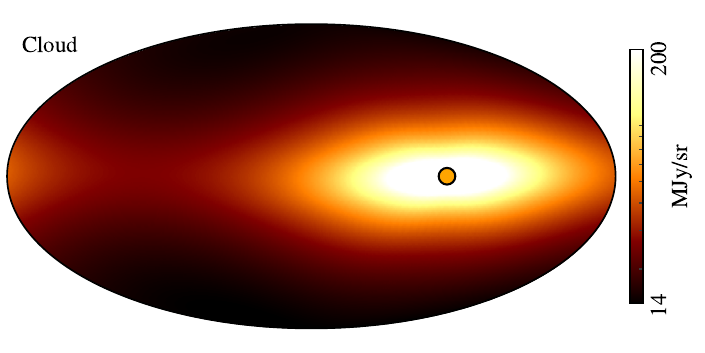}
	\includegraphics[width=0.495\linewidth]{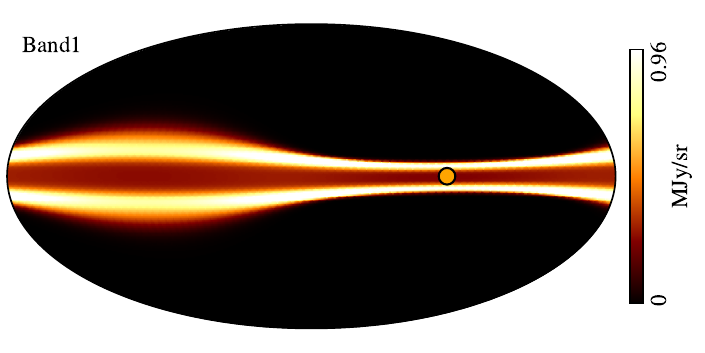}\\
	\includegraphics[width=0.495\linewidth]{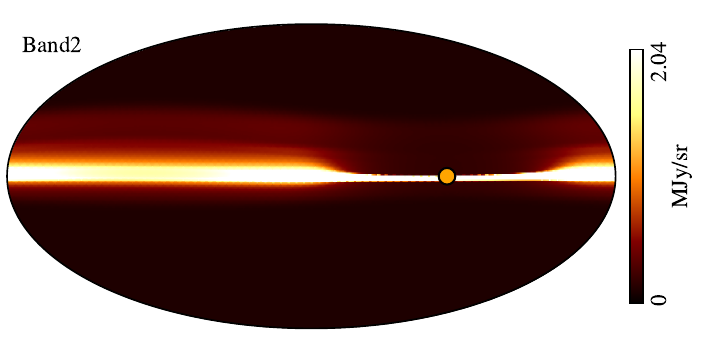}
	\includegraphics[width=0.495\linewidth]{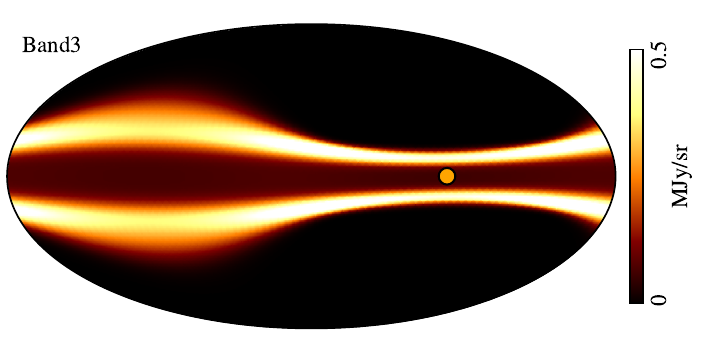}\\
	\includegraphics[width=0.495\linewidth]{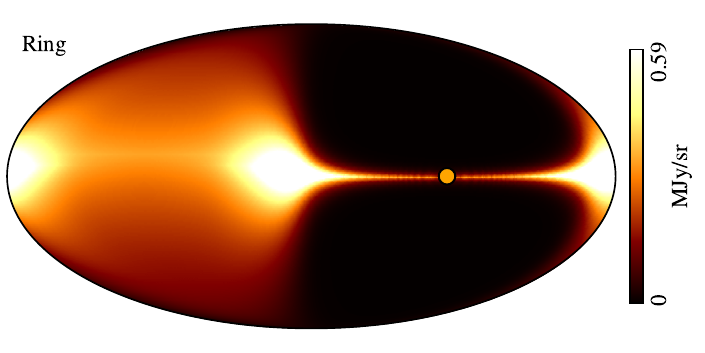}
	\includegraphics[width=0.495\linewidth]{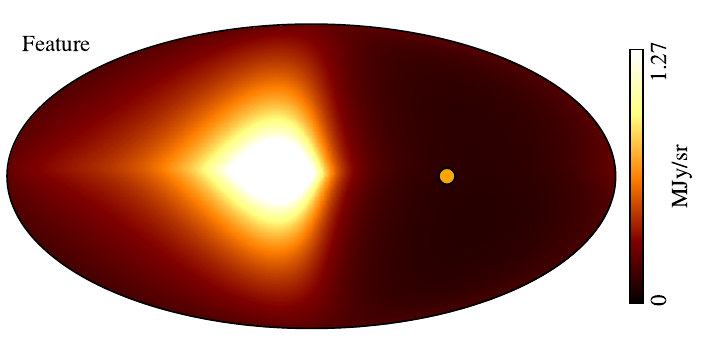}\\
	\caption{Instantaneous emission maps (January 1, 2020) for each IPD component in the DIRBE model. The Sun's location is marked with an orange circle. (\textit{Top left:}) The diffuse cloud. We note that the intensity is logarithmic due to the highly exponential nature of the zodiacal cloud density near the Sun. (\textit{Top right:}) Dust band 1. (\textit{Middle left:}) Dust band 2. (\textit{Middle right:}) Dust band 3. (\textit{Bottom left:}) The circumsolar ring. (\textit{Bottom right:}) The Earth-trailing feature.}
	\label{fig: inst_comps}
\end{figure*}
The emission can also be viewed component-wise, as illustrated in Fig. \ref{fig: inst_comps}, which shows the emission from Fig.~\ref{fig: inst_ecl_gal} split into the respective model components in ecliptic coordinates. The structure observed in these maps can be better understood by inspecting them alongside the density distributions seen in Figs.~\ref{fig: denscloudringfeature} and \ref{fig: densbands}. The diffuse cloud is the brightest IPD component at this wavelength, and as such, it looks very similar to the total ZE emission. We see that dust bands 1 and 3 appear wider on the side of the map facing away from the Sun. This widening is due to more high-latitude line of sights looking through the bright parts of the bands in this direction. The subtle monopole present in the map of dust band 2 results from the observer being encapsulated by dust, which we can see from the density distribution in the middle panel of Fig. \ref{fig: densbands}. The asymmetry of the emission with respect to the ecliptic is due to the inclination of the bands. 

The circumsolar ring and the Earth-trailing feature exhibit perhaps the most distinct structures in the maps. The bright stripe in the ecliptic wrapping behind the Sun in the map of the circumsolar ring (bottom left of Fig. \ref{fig: inst_ecl_gal}) can be understood as the observer looking past the Sun and through the opposite side of the circumsolar ring. The two bright regions symmetrically around the Sun in the ecliptic represent the observer looking tangential to Earth's orbit along the longest sections of the ring. We can understand the partial sky brightening in the circumsolar ring emission by noting that the observer is located at the inner edge of the ring (see the gray circle in the middle-right panel of Fig. \ref{fig: denscloudringfeature}).
This means that all directions pointing away from the Sun look through some portion of the circumsolar ring. A similar signature is observed in the Earth-trailing feature map (bottom-right of Fig. \ref{fig: inst_ecl_gal}; however, in this case, only one such bright region appears on the side pointing toward the feature. Since the Earth-trailing feature is much wider than the circumsolar ring in the $xz$ plane, the observer is again encapsulated by the dust in this component which results in another monopole. Satellite missions that observe from the second Sun-Earth Lagrange point (L2, $R\approx 1.01$~AU) should expect a slightly higher contribution from the circumsolar ring and Earth-trailing feature due to the observer being located closer to the denser regions of these components.

\section{Comparison with DIRBE} \label{sec: DIRBE}
In the following section, we illustrate a few use-cases of ZodiPy and validate the code implementation by applying the software to produce simulated timestreams using the pointing of DIRBE calibrated individual observations (CIO).\footnote{The DIRBE CIO is a user-friendly version of the raw time-ordered data observed by DIRBE. They are publicly available on LAMBDA:  \url{https://lambda.gsfc.nasa.gov/product/cobe/c_calib_i_o.html}.} We show that ZodiPy can accurately reproduce the ZE simulations created with the \textrm{COBE}-DIRBE Zodiacal Light Prediction Software used by the DIRBE team in their analysis of the experiment. Additionally, we use ZodiPy to produce binned simulated maps of the ZE at 25 $\mu$m as seen by DIRBE (using the CIO pointing) and compare these to binned maps of the DIRBE CIO. 
\subsection{Timestreams}
\begin{figure}
    \centering
    \includegraphics[width=\linewidth]{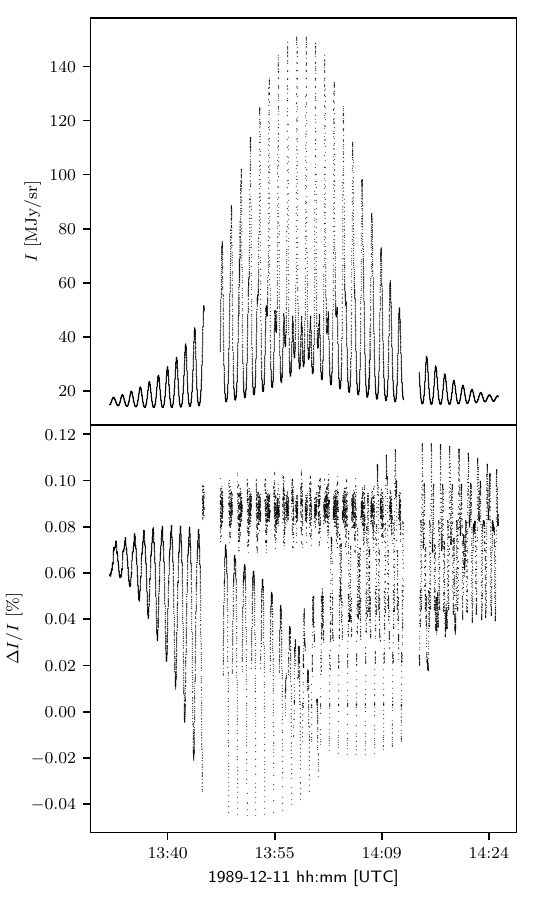}
    \caption{Zodiacal emission timestreams. (\textit{Top:}) Timestream simulated with ZodiPy of the ZE as predicted to have been observed by DIRBEs band 6 detector A on December 11, 1989 between $\sim$13:30 and $\sim$14:30 UTC. (\textit{Bottom:}) Error in percentage between the ZodiPy timestream in the top panel and a corresponding timestream produced using the DIRBE Zodiacal Light Prediction Software.}
    \label{fig:tods}
\end{figure}
The user input required to produce a timestream with the DIRBE Zodiacal Light Prediction Software is 1) a wavelength corresponding to one of the ten central band wavelengths of DIRBE; 2) the day of observation represented as an integer counting the number of days since January 1, 1990; 3) angular pointing information; and 4) an optional flag for whether or not to color-correct the output. In comparison, ZodiPy expects the following input arguments: 1) any wavelength or frequency; 2) the name of the observer (defined below) or explicit observer position in units of AU; 3) the time of observation; and 4) pointing information in the form of angles or HEALPix pixel indices. Since ZodiPy can produce timestreams for the arbitrary observer, the observer position is required explicitly. This argument is provided either by specifying the name of the observer, for which ZodiPy computes all required Solar System positions (given the time of observation) using the Astropy Solar System Ephemerides \citep{ASTROPY1, ASTROPY2} or by explicitly passing in the heliocentric Cartesian coordinates of the observer in units of AU. In order to compare the two codes, we need to consider the location of DIRBE explicitly.
The \textrm{COBE} satellite carrying the DIRBE instrument made observations from a near-Earth polar orbit at a distance of about 900 km above the surface. The variation in DIRBE's position from its orbit is insignificant compared to the scale at which the IPD is distributed. As such, we approximate DIRBE's location to be the center of Earth. 

The comparison of both codes is shown in Fig. \ref{fig:tods} where we have generated timestreams over a small subsequence of the pointing from DIRBE's first day of observations. The timestreams are simulated at 25 $\mu$m, which corresponds to the central wavelength of DIRBE's band 6, which is the band that observed the most ZE. The top panel shows the ZE timestream as predicted by ZodiPy, and the bottom panel shows the difference in percentage between the ZodiPy and DIRBE software predictions. We find the difference between the two codes to be less than 0.2\%. A similar comparison was also conducted for the other DIRBE bands, with results consistent with those presented here. We believe that the minor difference in the predictions made by the two codes may be attributed to one of or a combination of the following: 1) a difference in how the input pointing is mapped to unit vectors on the sphere; 2) numerical integration performance for evaluation of the line of sights; 3) a difference in the computed Solar System position; and 4) implementation of the Planck function.

\subsection{Binned emission maps}
\begin{figure*}
  \centering
	\includegraphics[width=0.45\linewidth, trim={0 0 1.5cm 0}, clip]{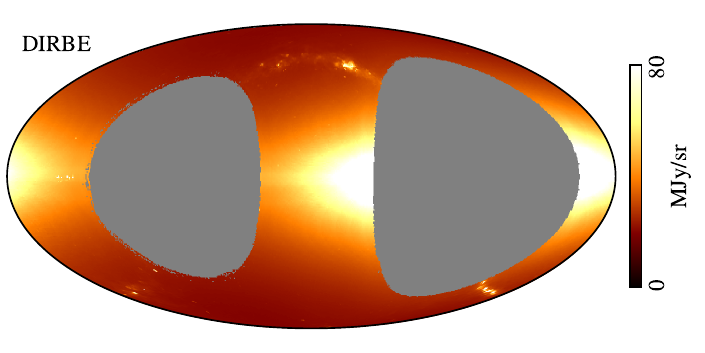}\hspace{0.01\linewidth}
	\includegraphics[width=0.52\linewidth]{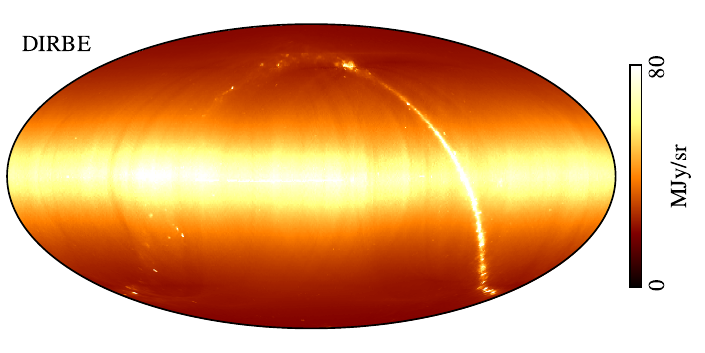}\\ \vspace{-0.3cm}
	\includegraphics[width=0.45\linewidth, trim={0 0 1.5cm 0}, clip]{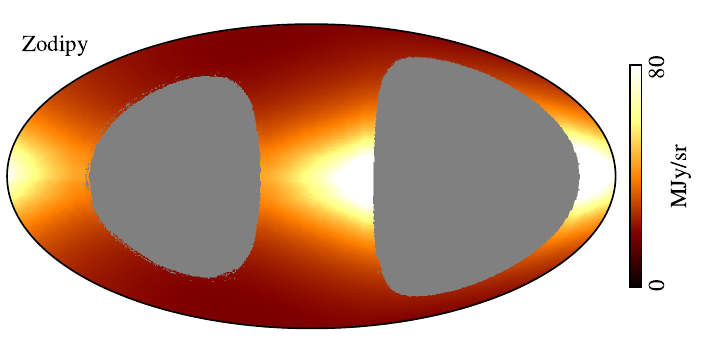}\hspace{0.01\linewidth}
	\includegraphics[width=0.52\linewidth]{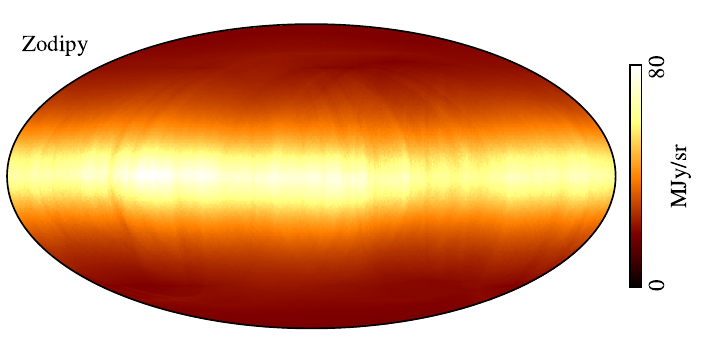}\\ \vspace{-0.3cm}
	\includegraphics[width=0.45\linewidth, trim={0 0 1.5cm 0}, clip]{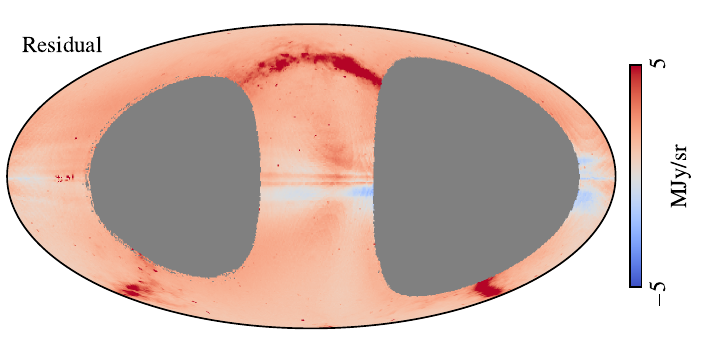}\hspace{0.01\linewidth}
	\includegraphics[width=0.52\linewidth]{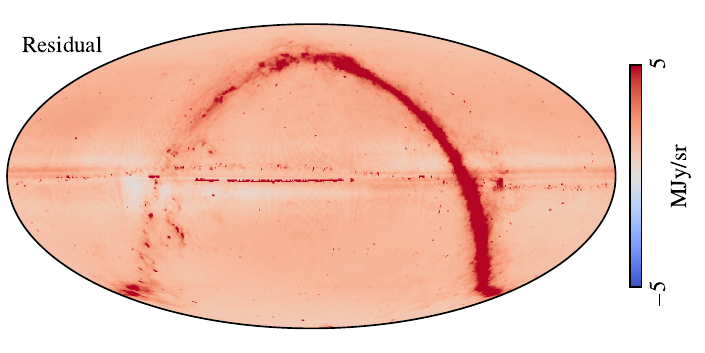}\\ \vspace{-0.3cm}
	\caption{(\textit{Top:}) Binned week maps of the DIRBE CIO. (\textit{Middle:}) Binned week maps from the ZodiPy simulations. (\textit{Bottom:}) DIRBE minus ZodiPy binned week map residuals. The columns correspond to week 1 (\textit{left}) and the full-survey (\textit{right}), respectively.
            }
	\label{fig: binned}
\end{figure*}
Figure~\ref{fig: binned} shows two instances of ZE corrections applied to the binned DIRBE CIO, which are comparable with Figs. 2c and 7 in K98. In the left column, we see the corrections applied to the first week of observations, and in the right column, we see the corrections applied to the full mission.

The top, middle, and bottom panels show the binned CIO, the corresponding ZodiPy predictions, and the CIO minus the ZodiPy predictions, respectively. We note the stripes and discontinuities in the full-survey maps. These are artifacts of the binning process and not features of the infrared sky. More specifically, these stripes correspond to differences in the path through the IPD distribution, as discussed in Sect. \ref{sec: time-varying}, where adjacent positions along the scanning pattern on the sky have been observed over extended time intervals.

We note that the simulations in this comparison were produced using a  modified version of the code that included the color-correction factor in Eq.~(1) in K98. This was done to make the simulated data more comparable to the CIO, which are color-corrected. The presented residual maps exhibit some weak digitization-like structures. We believe that this pattern is due to errors related to the conversion of the DIRBE pixel indices (natively given in the COBE quadrilateralized spherical cube format) to longitude and latitude rather than an inherent property of the CIO or the simulated emission.

\section{Conclusions}
We have presented software for modeling the thermal emission and scattered sunlight by IPD, as defined by the DIRBE and \Planck\ models, for arbitrary Solar System observers at both infrared wavelengths and subterahertz frequencies. The software is open-source and available on the Python Packaging Index. The code is designed to allow for new IPD model additions and updates with new developments to state-of-the-art models. Finally, as a validation of the code implementation, we have demonstrated that the software can accurately reproduce the Zodiacal corrections applied in the official analysis of the DIRBE data.

Currently, ZodiPy only models ZE in intensity. Polarization from thermal emission from ZE has never been detected, and it is expected to be less than 1\,\% \citep{weinberg1980, onaka2000, ganga21}. The scattered sunlight is known to be polarized as a function of the scanning strategy with the scattering angle and elongation, as shown in \citet{berriman94} and \citet{takimoto22}, for example. Allowing the user to predict the polarized emission at wavelengths where the ZE is dominated by scattered sunlight (1--5~$\mu$m) would help with forecasting for future polarized experiments. For a more complete discussion of the intrinsic polarized emission of ZE, readers can refer to \citet{ganga21} and references therein.

ZodiPy is an integral part of the \textsc{Cosmoglobe} framework. Part of \Cosmoglobe's goal is to perform end-to-end analysis using all microwave and infrared datasets to constrain the properties of the sky and the large-scale properties of the Universe. This code will be an invaluable tool for characterizing the ZE in archival datasets, such as \textrm{COBE}/DIRBE and \textit{Planck}/HFI, as well as future data from \textrm{JWST}, \textrm{SPHEREx}, and \textrm{LiteBIRD}. A joint analysis will not only help to better decontaminate data with ZE emission, but the joint analysis approach taken by \Cosmoglobe\ will necessarily improve the IPD model from K98. In combination with the full Gibbs sampling framework with \texttt{Commander3}, ZodiPy will be used to develop new models and visualize the individual components in an intuitive manner, all while providing the astronomical community at large with a simple interface to predict ZE in their own data.

\begin{acknowledgements}
  We thank Profs.~Hans Kristian Eriksen, Ken Ganga, and Ingunn Wehus for useful comments. We also thank the referee for their useful comments that improved this work. The current work has received funding from the European
  Union’s Horizon 2020 research and innovation programme under grant
  agreement numbers 819478 (ERC; \textsc{Cosmoglobe}) and 772253 (ERC;
  \textsc{bits2cosmology}). Some of the results in this paper have been derived using the HEALPix \citep{HEALPIX} package.
  We acknowledge the use of the Legacy Archive for Microwave Background Data
  Analysis (LAMBDA), part of the High Energy Astrophysics Science Archive Center
  (HEASARC). HEASARC/LAMBDA is a service of the Astrophysics Science Division at
  the NASA Goddard Space Flight Center.  
\end{acknowledgements}

%

\bibliographystyle{aa}
\bibliography{references,BP_bibliography}
\end{document}